\documentclass[11pt,a4paper]{article}
\usepackage[colorlinks=true, linkcolor=black!50!blue, urlcolor=blue, citecolor=blue, anchorcolor=blue]{hyperref}
\usepackage[font=small,labelfont=bf,margin=0mm,labelsep=period,tableposition=top]{caption}
\usepackage[a4paper,top=3cm,bottom=2.5cm,left=2.5cm,right=2.5cm,bindingoffset=0mm]{geometry}

\usepackage{graphicx,placeins}
\usepackage{float}
\usepackage{afterpage}
\usepackage{epsfig,cite,xspace}
\usepackage{amssymb}
\usepackage{amsmath}
\usepackage{dsfont}
\usepackage{multirow}
\usepackage{url}
\usepackage{xcolor,colortbl}
\usepackage{float}
\usepackage{afterpage}
\usepackage{url}
\usepackage{hyperref}
\usepackage{booktabs}
\usepackage{mathrsfs}

\usepackage{tikz}
\usepackage{tikz-3dplot}
\usepackage[compat=1.0.0]{tikz-feynman}

\usepackage{enumitem}
\usepackage{hyperref}
\usepackage{cite}

\usepackage{pifont}

\usetikzlibrary{shapes, arrows}
\usetikzlibrary{decorations.pathreplacing}
\usetikzlibrary{positioning, calc}
\tikzstyle{fitted} = [rectangle, minimum width=5cm, minimum height=1cm, text centered, draw=black, fill=red!30]
\tikzstyle{operations} = [rectangle, rounded corners, minimum width=2cm,text centered, draw=black, fill=red!30]
\tikzstyle{roundtext} = [rectangle, rounded corners, minimum width=2cm, minimum height=0.8cm, text centered, draw=black, fill=red!30]
\tikzstyle{n3py} = [rectangle, rounded corners, minimum width=3cm, minimum height=1cm, text centered, draw=black, fill=green!30]
\tikzstyle{myarrow} = [thick,->,>=stealth]
\tikzstyle{line} =[draw, -latex']
\tikzstyle{decision} = [diamond, draw, fill=red!20, text width=7.5em, text centered,  inner sep=0pt, minimum height=2em, aspect=4]
\tikzstyle{cloud} = [draw, ellipse,fill=green!20, minimum height=2em]
\tikzstyle{inout} = [rectangle, draw, fill=green!20, text width=9.5em, text centered, rounded corners, minimum height=2em, minimum width=10em]
\tikzstyle{block}=[rectangle, draw, fill=blue!20, text width=9.5em, 
                   text centered, rounded corners, minimum height=2em, 
                   minimum width=10em]

\definecolor{darkgreen}{rgb}{0.0, 0.5, 0.13}

\newcommand{\be}{\begin{equation}}

\newcommand{\ee}{\end{equation}}
\newcommand{\bea}{\begin{eqnarray}}
\newcommand{\eea}{\end{eqnarray}}
\newcommand{\bi}{\begin{itemize}}
\newcommand{\ei}{\end{itemize}}
\newcommand{\ben}{\begin{enumerate}}
\newcommand{\een}{\end{enumerate}}

\newcommand{\lc}{\left[}
\newcommand{\rc}{\right]}
\newcommand{\lp}{\left(}
\newcommand{\rp}{\right)}

\def\gsim{\mathrel{\rlap{\lower4pt\hbox{\hskip1pt$\sim$}}
    \raise1pt\hbox{$>$}}}         
\def\lsim{\mathrel{\rlap{\lower4pt\hbox{\hskip1pt$\sim$}}
    \raise1pt\hbox{$<$}}}         

\newcommand{\draft}[1]{}

\def\beq{\begin{equation}}
\def\eeq{\end{equation}}

\def\gsim{\mathrel{\rlap{\lower4pt\hbox{\hskip1pt$\sim$}}
    \raise1pt\hbox{$>$}}}         
\def\lsim{\mathrel{\rlap{\lower4pt\hbox{\hskip1pt$\sim$}}
    \raise1pt\hbox{$<$}}}         

\def\lapprox{\lower .7ex\hbox{$\;\stackrel{\textstyle <}{\sim}\;$}}
\def\gapprox{\lower .7ex\hbox{$\;\stackrel{\textstyle >}{\sim}\;$}}

\graphicspath{{figures/}}

\usepackage{tabularx}
\newcolumntype{C}[1]{>{\centering\arraybackslash}p{#1}}

\usepackage{amsmath}
\usepackage{amsfonts}
\usepackage{amssymb}
\usepackage{dsfont}
\usepackage{pifont}
\usepackage{booktabs}
\usepackage{graphicx}
\usepackage{epstopdf}
\usepackage{epsfig}
\usepackage{framed}
\usepackage{makeidx}
\usepackage{siunitx}
\usepackage[capitalise]{cleveref}
\usepackage{hyperref}
\usepackage{placeins}
\usepackage[font=small,labelfont=bf]{caption}

\RequirePackage{csquotes}
\usepackage{bbm}

\begin{document}
\newgeometry{top=1.5cm,bottom=1.5cm,left=1.5cm,right=1.5cm,bindingoffset=0mm}

\vspace{-2.0cm}
\begin{flushright}
    CERN-TH-2024-080\\TIF-UNIMI-2024-4
\end{flushright}
\vspace{0.3cm}

\begin{center}
  {\Large \bf LO, NLO, and NNLO Parton Distributions for LHC Event Generators}
  \vspace{1.1cm}

  Juan Cruz-Martinez$^1$, Stefano Forte$^2$, Niccolò Laurenti$^2$, Tanjona R. Rabemananjara$^{3,4}$,  and Juan Rojo$^{3,4}$   
    \vspace{0.7cm}
    
    {\it \small
      ~$^1$CERN, Theoretical Physics Department, CH-1211 Geneva 23, Switzerland\\[0.1cm]
    ~$^2$Tif Lab, Dipartimento di Fisica, Universit\`a di Milano and\\
      INFN, Sezione di Milano, Via Celoria 16, I-20133 Milano, Italy\\[0.1cm]
       ~$^3$Department of Physics and Astronomy, Vrije Universiteit, NL-1081 HV Amsterdam\\[0.1cm]
      ~$^4$Nikhef Theory Group, Science Park 105, 1098 XG Amsterdam, The Netherlands\\[0.1cm]
      }

    \vspace{2cm}

{\bf \large Abstract}

\end{center}

We present NNPDF4.0MC, a  variant of the NNPDF4.0 set of parton distributions
(PDFs) at LO, 
NLO and NNLO, with and without inclusion of the photon PDF, suitable
for use with  
Monte Carlo (MC) event generators, which require PDFs to satisfy
additional constraints in comparison to standard PDF sets. These
 requirements include PDF positivity  down to a low scale  $Q\sim 1$
 GeV, smooth extrapolation in the very small and
 large $x$ regions, and
 numerically stable results even in extreme regions of  phase
 space for all PDFs.
We compare the NNPDF4.0MC PDFs to their baseline NNPDF4.0
counterparts,
and to the 
NNPDF2.3LO set entering the {\sc\small Monash} tune of
the {\sc\small Pythia8} event generator.
We briefly assess the  phenomenological impact of these PDFs on  the
cross-sections for  hard and soft QCD processes at the LHC. 

\clearpage

\tableofcontents

\section{Introduction}
\label{sec:intro}

Monte Carlo (MC) event generators~\cite{Buckley:2011ms,Nason:2012pr,Hoche:2014rga,Campbell:2022qmc} provide a complete description of the final state in
high-energy particle collisions, and, as such, are an essential
ingredient in the interpretation of particle physics experiments. 
Widely used event generators for LHC physics include {\sc\small
  Pythia8}~\cite{Sjostrand:2007gs,Bierlich:2022pfr}, {\sc\small
  HERWIG7}~\cite{Bellm:2015jjp,Bellm:2019zci}, {\sc\small
  SHERPA}~\cite{Gleisberg:2008ta,Sherpa:2019gpd}, {\sc\small
  POWHEG}~\cite{Alioli:2010xd},  {\sc\small
  mg5\_aMC@NLO}~\cite{Alwall:2014hca}, and more recently
{\sc\small PanScales}~\cite{vanBeekveld:2023ivn,vanBeekveld:2022ukn,vanBeekveld:2022zhl,vanBeekveld:2024wws}.

Within a MC event generator, parton distributions
(PDFs)~\cite{Gao:2017yyd,Kovarik:2019xvh} are used not only in the
 evaluation of hadronic cross-section through their convolution with partonic
matrix elements, but also for the initial-state backwards parton
shower, and as  inputs to the modeling  of
non-perturbative phenomena~\cite{Skands:2010ak} such as the
underlying event (UE), multiple parton interactions (MPI), and related
soft QCD processes. 
For these latter aspects, PDFs should respect some  additional
constraints in comparison to default PDFs.
First, their usage in initial-state showers requires that they be
non-negative  down to the perturbative cutoff of $Q \simeq 1$ GeV.
Furthermore, their application to models of the UE, MPI, and other
low-energy QCD phenomena demands a very smooth extrapolation down to
very small $x$ and very small $Q^2$ values, and a gluon PDF that grows
sufficiently fast in the small $x$ region.
In order to prevent numerical problems associated to Monte Carlo
integration and sampling, PDFs should be numerically stable even in
extreme regions of  phase space which may be irrelevant for phenomenology.
Finally, in order to match to standard parton showers, the charm PDF
must be generated perturbatively (i.e.  an intrinsic component is not
allowed), and in order to account for electroweak corrections, the
possibility of including a photon PDF $\gamma(x,Q^2)$ 
and QED splittings in perturbative evolution should be allowed.

Several
groups~\cite{Lai:2009ne,Yan:2022pzl,Ball:2011uy,Sherstnev:2008dm,Sherstnev:2007nd}
have presented variants of their LO PDF sets, aimed to  usage in
MC event generators. 
For instance, the NNPDF2.3QED LO PDFs developed in~\cite{Ball:2012cx,Ball:2013hta,Carrazza:2013axa} were integrated in {\sc\small Pythia8}, and used as one of the inputs for its popular
{\sc\small Monash} tune~\cite{Skands:2014pea} of non-perturbative QCD physics.
Beyond LO, BFKL-resummed variants of the  NNPDF3.1 PDF set including the
constraints on the small-$x$ gluon  from $D$-meson production at LHCb
presented in~\cite{Ball:2017otu,Bertone:2018dse,Bertone:2017bme} also
satisfy the above requirements, and are available in {\sc\small
  Pythia8} as a stand-alone PDF set. 

Here we present variants of  NNPDF4.0~\cite{Ball:2021leu, NNPDF:2021uiq,NNPDF:2024djq,NNPDF:2024nan} at LO and, for the first time, NLO and NNLO,
tailored to their usage in modern MC event generators.
The main goal of these NNPDF4.0MC sets is to satisfy the
requirements discussed above, while at the same time providing the
best possible  
description of the NNPDF4.0 dataset, in particular at  NLO and NNLO.

\section{Methodology}
\label{sec:methodology}

Unless otherwise specified, we adopt the same experimental dataset,
theory calculations, and  methodology used in the construction of
the recent MHOU, QED, and
aN$^3$LO NNPDF4.0 PDF sets~\cite{NNPDF:2024djq,NNPDF:2024nan,NNPDF:2024dpb}.
In particular, we exploit the new NNPDF theory
pipeline~\cite{Barontini:2023vmr} built upon the {\sc\small
  EKO}~\cite{Candido:2022tld} evolution code, {\sc\small YADISM} DIS
module~\cite{Candido:2024rkr}, and {\sc\small PineAPPL} fast grid interface~\cite{Carrazza:2020gss}.
The same values of the input SM parameters are used,  in particular $\alpha_s(m_Z)=0.118$ for the LO, NLO, and NNLO fits.
We only provide a central PDF, instead of a set of PDF replicas
representing the PDF probability distribution, because in the presence of extra
constraints uncertainties might become unreliable, and they are anyway
not relevant for  applications to  MC event generators.

\paragraph{Positivity and perturbative charm.}
Positivity of MC PDFs is required both for their usage in the initial-state shower as well as for the modeling of soft QCD phenomena.
At LO, PDFs can be identified with physical cross-sections and hence are positive-definite.
This is not necessarily true at NLO and beyond, where PDFs become scheme dependent and may be negative in certain regions of the phase space.
Whereas in the commonly used $\overline{\rm MS}$ scheme PDFs are
positive also at NLO and beyond, this only holds in the perturbative
region, i.e. at high enough
scale~\cite{Candido:2020yat,Collins:2021vke,Candido:2023ujx}, and
correspondingly 
PDF positivity may  fail when extrapolating to low $Q$ values.  

In the baseline NNPDF4.0 analysis, PDF positivity is imposed at the
initial parametrization scale ($Q_0=1.65$ GeV) at LO and at a higher
scale, $Q^2_{\rm pos} = 5$ GeV$^2$, at NLO and beyond, following the prescription of~\cite{Candido:2020yat,Candido:2023ujx}.
In addition, positivity of a set of physical observables at $Q^2_{\rm pos}$ is also imposed.
Therefore, within the NNPDF4.0 methodology, the NLO and NNLO PDFs may be negative at low values of $Q^2$ as long as, upon evolution, they become positive at $Q^2 \geq Q^2_{\rm pos}$.
Even though this may happen in regions of phase space for which there
are no  direct experimental constraints, or such that a fixed-order
leading-twist approximation breaks down, positivity  is
nevertheless required by MC generators.
Furthermore, in the default NNPDF4.0 sets
the charm PDF is parametrized and determined from 
the data on the same footing as all other PDFs~\cite{Ball:2016neh},
with its behavior for $Q <Q_0$ determined 
by backwards QCD evolution together with the matching from the $n_f=4$
to the $n_f=3$ flavor scheme~\cite{Ball:2022qks}. However,
a variant of NNPDF4.0 in which charm vanishes in the  $n_f=3$ flavor
scheme and is determined by perturbative matching conditions  in the  $n_f=4$
scheme is also available;  in this case PDFs are parametrized at
$Q_0=1$ GeV, hence below the matching scale,  set at $\mu_c=m_c=1.51$ GeV.

We consequently start from this  perturbative charm variant of NNPDF4.0,
with   perturbative matching
conditions used to determine charm at the  matching scale $\mu_c=m_c$.
We
then impose the positivity of $g(x,Q_0)$ and $\Sigma(x,Q_0)$ at
$Q_0=1$ GeV by squaring the corresponding neural network outputs. 
This ensures positivity of the gluon and the quark singlet PDFs at
$Q_0=1$~GeV and  consequently also for $Q > Q_0$ thanks to their rise
at small $x$ induced by perturbative 
QCD evolution as
the scale is increased.
Positivity of individual quark and antiquark PDFs is imposed
at  $Q^2_{\rm pos} = 5$ GeV$^2$  as in the default. This is  sufficient to
guarantee positivity down to $Q_0$ both at large $x$, where
 perturbative evolution is
moderate even at low scale, and also at small $x$, where nonsinglet
PDFs vanish.  
This strategy leads to  positive-definite PDFs in the full range
of $(x,Q^2)$ 
probed by MC generators at LO and NLO. 

At NNLO, the perturbative
matching conditions lead to a charm PDF that at $Q=m_c$ is negative at
small $x\lesssim 10^{-2}$, though it is already positive at all $x$ for
$Q^2\gtrsim5$~GeV$^2$. Hence at NNLO it is not possible to
simultaneously satisfy at $\mu_c=m_c$
the requirements that charm be positive and
determined by perturbative matching.
As we will discuss in Sect.~\ref{sec:results}, the low-scale
positivity of the gluon at small $x$ is disfavored by the data and
consequently imposing it leads to some deterioration of the fit quality.

\paragraph{Extrapolation in $x$ and $Q^2$.}
General-purpose MC event generators should provide reliable results for the broadest possible region of phase space.
This requires input PDFs with a smooth behavior in a wide $Q$ range,
from $Q\simeq 1$ GeV (initial-state showers, non-perturbative QCD
modeling) up to $Q\sim 100$ TeV (relevant for future particle
colliders and for applications to astroparticle physics) and from
$x\simeq 10^{-9}$ (forward particle production) all the way up to
large-$x$ values close to the
elastic limit $x=1$ (required for high-mass new physics searches). 
Since these regions extend beyond the coverage of available data, a
robust extrapolation procedure is necessary. 

While PDF extrapolation in $Q^2$ is fixed by perturbative QCD
evolution, extrapolation in $x$ depends on  assumptions. 
In the NNPDF4.0 approach, extrapolation to the  small $x$
and large $x$ regions is provided by the output of a preprocessed neural
network, and thus controlled by the behavior of both the neural net
and the preprocessing function. 
This  extrapolation to low $Q^2$ and large $x$ values might be
affected by numerical
instabilities, both native,  and related to their storage as
{\sc\small LHAPDF} grids. 
Specifically,  the  low $Q^2$ behavior is controlled by evolution from
higher scales, that may amplify small differences in the initial
condition, due to the growing value of $\alpha_s(Q)$, while at large
$x$  PDFs become very small and thus particularly sensitive to numerical instabilities. 
These two issues are intertwined, since even small $\mathcal{O}\lp 10^{-5}\rp$ numerical
differences in the solution of evolution equations
 may be  enough to  distort the PDFs in the
large $x$ region where they are almost vanishing.
While such instabilities are innocuous for  phenomenological
applications,  they may lead to numerical issues when PDFs are used
in MC generators. 

In order to prevent these  instabilities and ensure that the MC PDFs are
everywhere smooth
and well-behaved,  the NNPDF4.0MC PDFs
are delivered as
an {\sc\small LHAPDF} grid with a finer coverage in $x$ for the region $x\in \lc 0.7, 0.95\rc $.
For $x \gtrsim 0.95$, PDFs essentially vanish and any residual
oscillations can be safely set to zero. 
In addition, instabilities of the order of the accuracy of the
{\sc\small LHAPDF} interpolation are averaged 
out by means of a dedicated Gaussian filter. Possible issues related
to backward evolution are prevented  by parametrizing PDFs at $Q_0=1$~GeV, so no backward evolution is needed.
We thus deliver {\sc\small LHAPDF} grids that provides an interpolated
output for all $x\in \lc 10^{-9},1\rc$
and $Q\in \lc 1, 10^6\rc$ GeV.

\paragraph{QED evolution and the photon PDF.}
As shown
in~\cite{Bertone:2017bme,NNPDF:2024djq,Manohar:2016nzj,Cridge:2021pxm,Xie:2021equ}
and related studies, the impact of inclusion of a photon PDF alongside
quark and gluon PDFs is moderate, its main effect being a reduction
of the gluon momentum fraction by up to around $0.5\%$ in favor of the photon.
Here we take the photon PDF $\gamma(x,Q^2)$ at $Q=1$~GeV from the
NNPDF4.0 QED NNLO PDF set~\cite{NNPDF:2024djq},  we
include it as boundary condition to the  QCD$\otimes$QED
evolution of the LO, NLO, and NNLO NNPDF4.0MC PDFs, and impose a
momentum sum rule that now also includes a photon contribution.
We adopt the so-called exact-iterated (EXA) solution of the
QCD$\otimes$QED  evolution equations, as implemented in {\sc\small
  EKO}~\cite{Candido:2022tld}, as in Ref.~\cite{NNPDF:2024djq} to
which we refer for more details. For pure QCD evolution we use instead the
 truncated (TRN) solution as in Ref.~\cite{Ball:2021leu}, so that in
 each case the PDF sets presented here are based on the same form of
 the solution of the evolution equations as their default counterparts.

\paragraph{NNPDF4.0MC overview.}
In
Table~\ref{table:fit_settings} we summarize the 
settings adopted for the NNPDF4.0MC PDFs,  compared to those of their
baseline counterparts: LHAPDF naming ID, publication reference,
PDF parametrization scale and solution of the evolution equations,
positivity scale, value of $\alpha_s(m_Z)$, and treatment of charm
(data-driven, or determined from perturbative matching).
In this table $q_i,\bar{q}_i$ denote light (up, down, and strange)
quarks and antiquark PDFs, as, following~\cite{Candido:2020yat,Candido:2023ujx},  positivity
of the charm PDF is never  imposed.

\begin{table}[htbp]
  \centering
  \scriptsize
   \renewcommand{\arraystretch}{2.2}
   \begin{tabularx}{\textwidth}
   {|X|c|c|c|c|c|c}
     \toprule
  ID  & Ref. & evolution ($Q_0$)    & Positivity ($Q_{\rm pos}$)  & $\alpha_s(m_Z)$  &   Charm     \\
  \midrule
      {\tt NNPDF23\_lo\_as\_0130\_qed}  & \cite{Carrazza:2013axa} &  QCD$_{\rm LO}\otimes$QED$_{\rm LO}$ TRN (1.0 GeV)   & $g,q_i,\bar{q}_i >0$ ($1$ GeV)  & 0.130 & pert.  \\
      \hline
  {\tt NNPDF40\_lo\_as\_01180}  & \cite{Ball:2021leu} &     QCD$_{\rm LO}$ TRN (1.65 GeV) & $g,q_i,\bar{q}_i >0$ ($1.65$ GeV)  & 0.118 & fitted  \\\hline
  {\tt NNPDF40\_lo\_pch\_as\_01180}  & \cite{Ball:2021leu} &     QCD$_{\rm LO}$ TRN (1.65 GeV) & $g,q_i,\bar{q}_i >0$ ($1$ GeV)  & 0.118 & pert.  \\\hline
  {\tt NNPDF40MC\_lo\_as\_01180}  & t.w. &    QCD$_{\rm LO}$ TRN (1.0 GeV)   & $g,q_i,\bar{q}_i >0$ ($1$ GeV)  & 0.118 & pert.  \\\hline
  {\tt NNPDF40MC\_lo\_as\_01180\_qed}  & t.w. &    QCD$_{\rm LO}\otimes$QED$_{\rm LO}$ EXA (1.0 GeV)   & $g,q_i,\bar{q}_i >0$ ($1$ GeV)  & 0.118 & pert.  \\
  \midrule
      {\tt NNPDF40\_nlo\_as\_01180}  & \cite{Ball:2021leu} &     QCD$_{\rm NLO}$ TRN (1.65 GeV) & $g,q_i,\bar{q}_i >0$ ($\sqrt{5}$ GeV)  & 0.118 & fitted  \\\hline
      {\tt NNPDF40\_nlo\_pch\_as\_01180}  & \cite{Ball:2021leu} &     QCD$_{\rm NLO}$ TRN (1 GeV) & $g,q_i,\bar{q}_i >0$ ($\sqrt{5}$ GeV)  & 0.118 & pert.  \\\hline
      \multirow{2}{*}{\tt NNPDF40MC\_nlo\_as\_01180}  & \multirow{2}{*}{t.w.} &     \multirow{2}{*}{QCD$_{\rm NLO}$ TRN (1 GeV)} & $g,\Sigma >0$ ($1$ GeV)  & \multirow{2}{*}{0.118} & \multirow{2}{*}{pert.}  \\
       &      &     &  $q_i,\bar{q}_i >0$ ($\sqrt{5}$ GeV)    &      &    \\\hline
               {\tt NNPDF40\_nlo\_as\_01180\_qed}  & \cite{NNPDF:2024djq} &  QCD$_{\rm NLO}\otimes$QED$_{\rm NLO}$    EXA (1.65 GeV) & $g,q_i,\bar{q}_i >0$ ($\sqrt{5}$ GeV)  & 0.118 & fitted  \\\hline
                \multirow{2}{*}{\tt NNPDF40MC\_nlo\_as\_01180\_qed}  & \multirow{2}{*}{t.w.} &     \multirow{2}{*}{QCD$_{\rm NLO}\otimes$QED$_{\rm NLO}$ EXA (1 GeV)} & $g,\Sigma >0$ ($1$ GeV)  & \multirow{2}{*}{0.118} & \multirow{2}{*}{pert.}  \\
                &      &     &  $q_i,\bar{q}_i >0$ ($\sqrt{5}$ GeV)    &      &    \\
                \midrule
                   {\tt NNPDF40\_nnlo\_as\_01180}  & \cite{Ball:2021leu} &     QCD$_{\rm NNLO}$ TRN (1.65 GeV) & $g,q_i,\bar{q}_i >0$ ($\sqrt{5}$ GeV)  & 0.118 & fitted  \\\hline
      {\tt NNPDF40\_nnlo\_pch\_as\_01180}  & \cite{Ball:2021leu} &     QCD$_{\rm NNLO}$ TRN (1 GeV) & $g,q_i,\bar{q}_i >0$ ($\sqrt{5}$ GeV)  & 0.118 & pert.  \\\hline
      \multirow{2}{*}{\tt NNPDF40MC\_nnlo\_as\_01180}  & \multirow{2}{*}{t.w.} &     \multirow{2}{*}{QCD$_{\rm NNLO}$ TRN (1 GeV)} & $g,\Sigma >0$ ($1$ GeV)  & \multirow{2}{*}{0.118} & \multirow{2}{*}{pert.}  \\
         &      &     &  $q_i,\bar{q}_i >0$ ($\sqrt{5}$ GeV)    &      &    \\ \hline
               {\tt NNPDF40\_nnlo\_as\_01180\_qed}  & \cite{NNPDF:2024djq} &  QCD$_{\rm NNLO}\otimes$QED$_{\rm NLO}$    EXA (1.65 GeV) & $g,q_i,\bar{q}_i >0$ ($\sqrt{5}$ GeV)  & 0.118 & fitted  \\\hline
                \multirow{2}{*}{\tt NNPDF40MC\_nnlo\_as\_01180\_qed}  & \multirow{2}{*}{t.w.} &     \multirow{2}{*}{QCD$_{\rm NNLO}\otimes$QED$_{\rm NLO}$ EXA (1 GeV)} & $g,\Sigma >0$ ($1$ GeV)  & \multirow{2}{*}{0.118} & \multirow{2}{*}{pert.}  \\
                &      &     &  $q_i,\bar{q}_i >0$ ($\sqrt{5}$ GeV)    &      &    \\
 \bottomrule
   \end{tabularx}
   \vspace{0.3cm}
   \caption{\small The NNPDF4.0MC PDFs presented in this work (t.w.)
     and their baseline counterparts. 
\label{table:fit_settings}
}
\end{table}


\section{The NNPDF4.0MC PDFs}
\label{sec:results}

\begin{figure}[!tb]
  \includegraphics[width=0.99\textwidth]{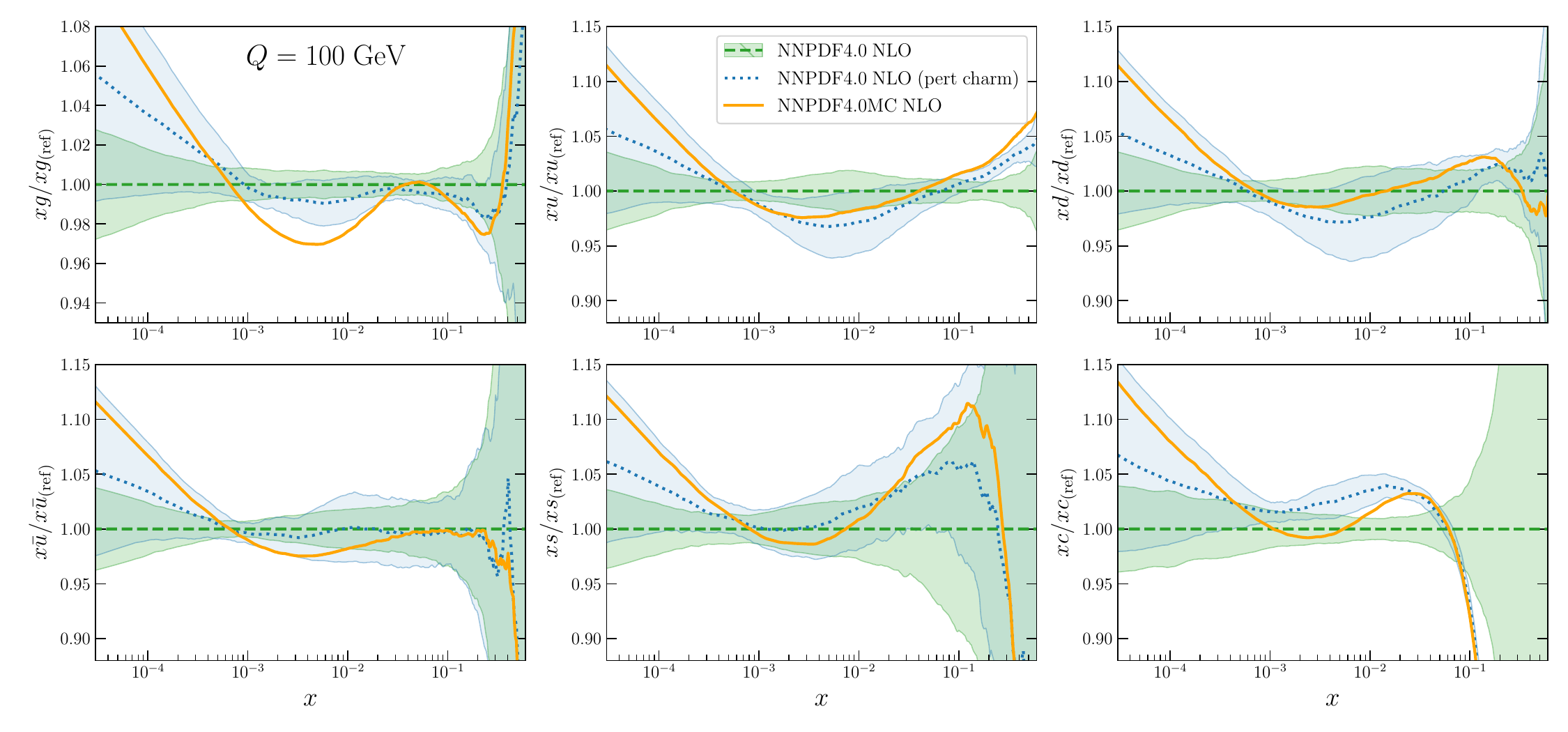}
  \caption{The NLO NNPDF4.0MC gluon, up, down, antiup, strange and
    charm PDFs  at $Q=100$~GeV (from left to right and from top to bottom),  shown as a ratio to their baseline counterpart. The
    uncertainty shown is the 68\% CL on the baseline. The baseline
    variant with perturbative charm is also shown.
  }
  \label{fig:xdep-q100-impact-MCPDF-constraints}
\end{figure}

We now compare the NNPDF4.0MC PDF sets to the baseline NNPDF4.0 fits and to  the NNPDF2.3QED LO PDFs used for  the {\sc\small
  Monash} tune~\cite{Skands:2014pea} of {\sc\small Pythia8}. 
Here we only present some  representative results; an extensive set of
comparisons is available
online.\footnote{\url{https://data.nnpdf.science/vp-public/NNPDF40MC_comparisons/}}
In all comparisons below, unless otherwise stated, NNPDF4.0 refers to
the default sets, and indeed the purpose of the comparison is to
illustrate the difference in phenomenology to be expected if the MC
sets instead of the default are used, for instance in applications to
experimental analysis. In particular, a comparison 
to the 
perturbative charm variants of NNPDF4.0 listed in
Table~\ref{table:fit_settings} will only be shown in
Figs.~\ref{fig:xdep-q100-impact-MCPDF-constraints}-\ref{fig:lumi-mcpdfs.pdf},
for the sake of assessing the impact of this particular assumption
among the others that characterize the NNPDF4.0MC sets.

The fit quality for the  NLO and NNLO  PDF sets
of Table~\ref{table:fit_settings}  is summarized in
Table~\ref{tab:chi2_TOTAL}, where we show 
the number of data points and the $\chi^2$ per data point; LO $\chi^2$
values
are not shown since fit quality at LO is generally poor and the
specific value of the $\chi^2$ is not significant. When comparing fit
quality, the MC PDFs constructed here should be viewed as PDFs that
include some additional theory assumptions: for instance, the positive
small
$x$ behavior of the gluon at low scale can be justified based on
non-perturbative physics arguments (see
e.g.~\cite{Sjostrand:2021dal}). Because extra constraints are
introduced, the agreement with the data
of the MC PDFs will be either unchanged, or possibly
worse than that of the default, i.e. the fit quality will deteriorate (or
remain unchanged). The purpose of the comparison is then
to check that the deterioration in fit quality is not such as to
rule out these extra assumptions.

For pure QCD PDFs, we find that at NLO (NNLO) the total $\chi^{2}$
per data point of the baseline fit increases from 1.28~(1.16) to
1.30~(1.22), 
an effect of about 1$\sigma$~(3$\sigma$) in units of the statistical variance of the $\chi^2$ distribution for $n_{\rm dat}= 4443$~(4626) data points.
Therefore, imposing the MC PDF conditions at NLO cannot be
distinguished from a change in $\chi^2$ value due to a random
fluctuation of the data. At NNLO the MC conditions do lead to a mild
deterioration of fit quality, related to the fact that the rapid rise
of the gluon at small $x$ as the scale increases tends to lead in turn to a
negative gluon at scales $Q^2\lesssim$~few
GeV$^2$~\cite{Candido:2023ujx}. This rise is stronger at NNLO, and at
low scale NNLO corrections become large; consequently at NNLO
a low-scale positive gluon is more difficult to accommodate, though again
it cannot be excluded. 
For the QCD$\otimes$QED sets, the same behavior is observed at NNLO,
while now at NLO a more significant deterioration of fit quality is seen.
This can be traced to the fact that  subleading terms
included in the EXA solution of the
evolution equations lead to perturbative evolution that is faster  than
for the TRN solution, especially 
when the anomalous dimension is large, which then makes the problem
with low-scale gluon positivity more serious at NLO. The difference
between the pure 
QCD and QCD$\otimes$QED cases at NLO should thus be viewed as driven by
missing NNLO QCD corrections.

\begin{table}[t]
  \small
  \centering
  \renewcommand{\arraystretch}{1.4}
  \begin{tabularx}{\textwidth}{X|r|cc|cc|r|cc|cc|}
   \toprule
      \multirow{3}{*}{Dataset by process group}
    & \multicolumn{5}{c|}{NLO}
    & \multicolumn{5}{c}{NNLO} \\
    & \multirow{2}{*}{$n_{\rm dat}$}
    & \multicolumn{2}{c}{QCD}
    & \multicolumn{2}{c|}{QCD+QED}
    & \multirow{2}{*}{$n_{\rm dat}$}
    & \multicolumn{2}{c}{QCD}
    & \multicolumn{2}{c}{QCD+QED}
  \\
                                  &      & BL   & MC   & BL   & MC   &      & BL   & MC   & BL   & MC \\ \midrule
  DIS NC                          & 1953 & 1.35 & 1.37 & 1.38 & 1.54 & 2110 & 1.22 & 1.30 & 1.22 & 1.29 \\
  DIS CC                          & 988  & 0.91 & 0.92 & 0.94 & 0.95 & 989  & 0.90 & 0.89 & 0.90 & 0.89 \\
  DY NC                           & 669  & 1.58 & 1.84 & 1.67 & 2.04 & 736  & 1.20 & 1.30 & 1.22 & 1.33 \\
  DY CC                           & 197  & 1.38 & 1.56 & 1.40 & 1.61 & 157  & 1.45 & 1.55 & 1.47 & 1.57 \\
  Top pairs                       & 66   & 2.40 & 2.14 & 2.51 & 2.47 & 64   & 1.27 & 1.16 & 1.31 & 1.27 \\
  Single-inclusive jets           & 356  & 0.82 & 0.88 & 0.83 & 0.93 & 356  & 0.94 & 1.01 & 0.93 & 1.00 \\
  Dijets                          & 144  & 1.51 & 1.55 & 1.56 & 1.62 & 144  & 2.01 & 2.01 & 1.94 & 1.93 \\
  Photon                          & 53   & 0.57 & 0.60 & 0.64 & 0.74 & 53   & 0.76 & 0.67 & 0.74 & 0.68 \\
  Single top                      & 17   & 0.36 & 0.36 & 0.38 & 0.36 & 17   & 0.37 & 0.38 & 0.39 & 0.40 \\ \midrule
  Total                           & 4443 & 1.28 & 1.30 & 1.30 & 1.44 & 4626 & 1.16 & 1.22 & 1.17 & 1.22
  \\
\bottomrule
\end{tabularx}
  \vspace{0.3cm}
  \caption{The number of data points and the $\chi^2$ per data point for the NLO and NNLO baseline NNPDF4.0 fits (BL), compared to their NNPDF4.0MC counterparts (MC), with 
    the same process categorisation as in Ref.~\cite{NNPDF:2024dpb}.
    The $\chi^2$ values are provided for the QCD-only
    (\texttt{NNPDF40(MC)\_<order>\_as\_01180}) and for the
    QCD$\otimes$QED (\texttt{NNPDF40(MC)\_<order>\_as\_01180\_qed})
    fits of Table~\ref{table:fit_settings}.
  }
  \label{tab:chi2_TOTAL}
\end{table}

\begin{figure}[!tb]
  \includegraphics[width=0.99\textwidth]{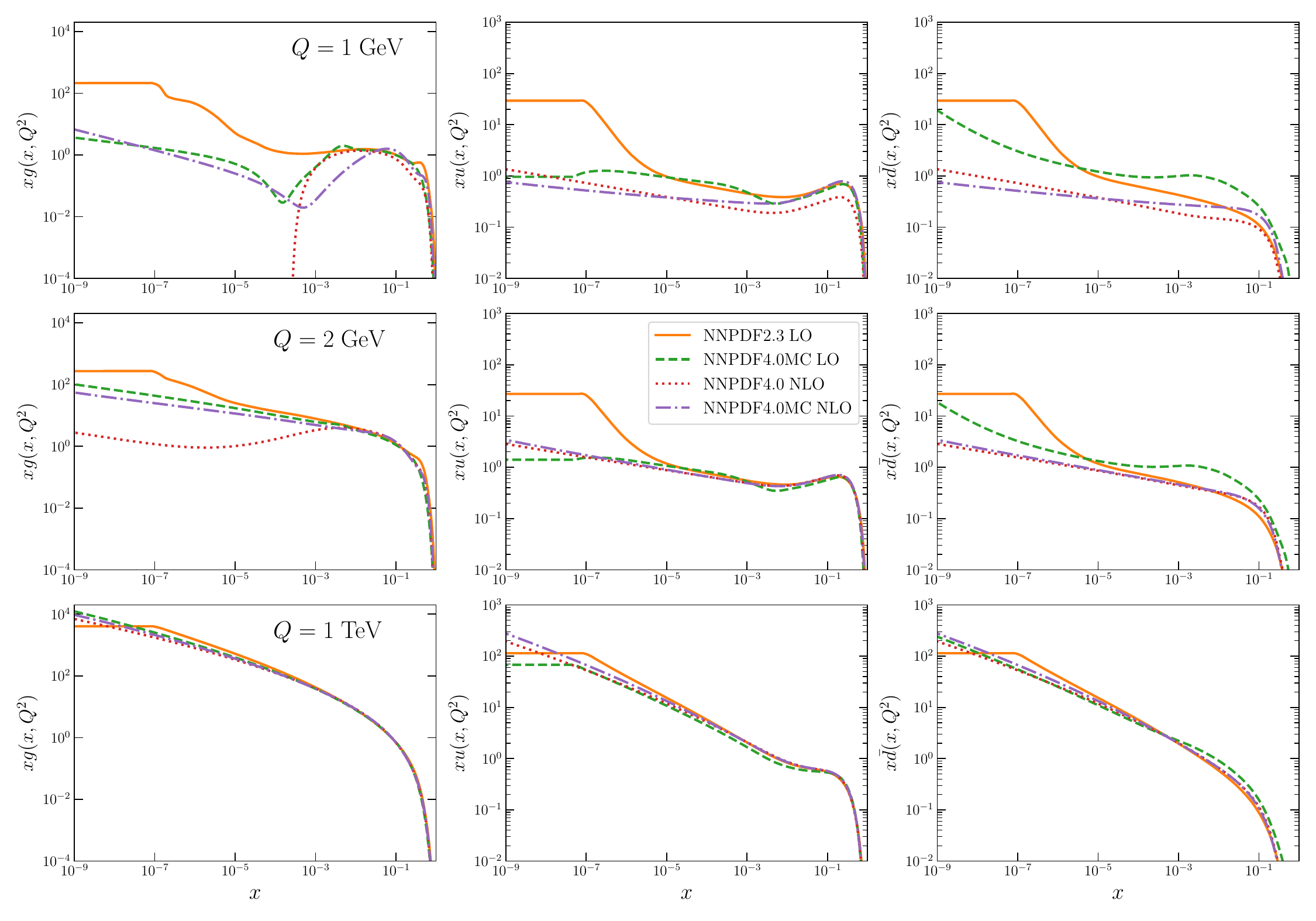}
  \caption{The NNPDF4.0MC LO and NLO gluon, up, and antidown PDFs
    (from left to right)
    compared to NNPDF2.3LO and NNPDF4.0 NLO, at three scales:
    $Q=1$~GeV, 2~GeV, and $1$~TeV (from top to bottom).
    Only central values are shown, in the region for which PDFs
    are provided via 
{\sc\small LHAPDF}.
  }
  \label{fig:xdep-q1geV-smallx}
\end{figure}

The MC and baseline LO and NLO PDFs are compared in
Fig.~\ref{fig:xdep-q1geV-smallx}, where we  display the gluon, up and
antidown PDFs at 
$Q=1$~GeV, $2$~GeV, and $1$ TeV. 
Recall that the small-$x$ behavior of
all quark and antiquark PDFs is the same, and dominated by that of the
singlet quark distribution. 
We show the full $x$ region in  which the NNPDF4.0MC PDFs are provided via the
{\sc\small LHAPDF} interpolation, i.e. $10^{-9}\le x \le 1$.
Note that
the NNPDF2.3LO set was only provided for $x\ge 10^{-7}$, while
for smaller $x$ values  PDFs are frozen to their
 value at $x=10^{-7}$. Apart from this trivial difference, the main
 difference between 
 the 2.3 and 4.0 LO sets is that for NNPDF4.0MC the
 rise of the small-$x$ gluon is qualitatively similar at LO and NLO, a
 feature facilitating the 
tuning of soft QCD models in MC event generators. This is due to the
greater theoretical consistency of assumptions between LO and NLO in
the NNPDF4.0MC sets, specifically the choice of the same value of
$\alpha_s$.   The main
 difference between the MC and default NLO PDFs is related to the small
 $x$ positivity of the gluon at low scale. 
As the scale $Q$ is increased, relative differences between the
various PDF sets are washed out by perturbative evolution.

\begin{figure}[!tb]
    \includegraphics[width=0.99\textwidth]{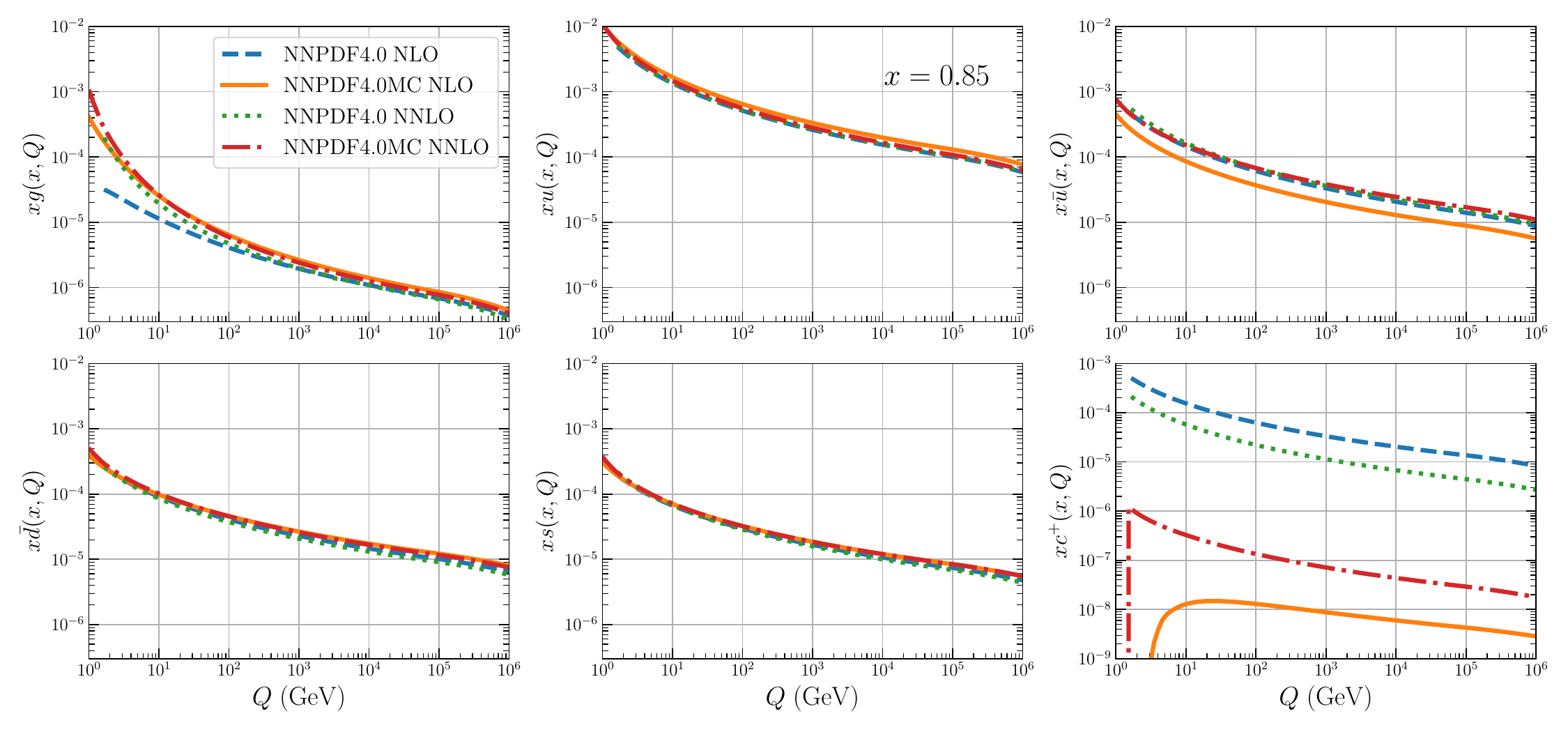}
    \caption{The NNPDF4.0MC NLO and NNLO gluon, up, antiup, antidown,
      strange and total charm PDFs
      (from left to right and from top to bottom), compared to their
      baseline counterparts as a function of scale for a fixed large 
$x=0.85$ value.
      The range $1\leq Q\leq 10^6$~GeV shown  corresponds to the full interpolation
      range provided by the {\sc\small LHAPDF} grids that we deliver.
    }
    \label{fig:largex-Qdep-extrapolation}
\end{figure}

In order to demonstrate smoothness of the NNPDF4.0MC sets in the
large-$x$ extrapolation region,  we display in
Fig.~\ref{fig:largex-Qdep-extrapolation}  the
NLO and NNLO NNPDF4.0MC PDFs for $x=0.85$ as a function of scale,
compared to the central value of their baseline counterparts. 
The $Q$ range shown corresponds to the full interpolation
range  in the {\sc\small LHAPDF} grids that we provide.
All PDFs displayed exhibit a satisfactory level of smoothness.

In order to fully assess the difference between the MC sets and their
baseline counterparts, in
Fig.~\ref{fig:xdep-q100-impact-MCPDF-constraints} we display the ratio
of the NNPDF4.0MC NLO PDFs to the baseline,  also showing  the 68\% CL
PDF uncertainties on the latter. 
In order to trace the  origin of
differences, the NNPDF4.0 NLO set with perturbative charm of
Table~\ref{table:fit_settings} is also shown.
In the region $x\gsim 10^{-3}$, where the bulk of experimental data is
located, the quark MC PDFs are mostly contained within the uncertainty
band of the baseline. Larger differences, that can be traced to the requirement of low-scale
positivity, are observed for the gluon
PDF, especially at small 
$x\lesssim 10^{-2}$. These in turn propagate onto the other PDFs at small $x$,
all of which display a stronger small-$x$ rise in comparison to the
baseline in the extrapolation region  $x\lesssim 10^{-3}$.

The results of Fig.~\ref{fig:xdep-q100-impact-MCPDF-constraints} imply that the additional model assumptions entering the MC PDFs
do not distort the baseline PDFs in the bulk of the data region beyond the $1\sigma$
level, indicating that most LHC cross-sections obtained with the
NNPDF4.0MC sets will be consistent with those derived using the
baseline PDFs. In fact, it is clear from
Fig.~\ref{fig:xdep-q100-impact-MCPDF-constraints} that for most PDFs, especially for
the sea quark PDFs, a large part of
the difference between the MC PDFs and the default is due to  having
adopted perturbative charm. 

\section{Impact on LHC physics}
\label{sec:pheno}

We now carry out a brief assessment of the phenomenological impact of
similarities and 
differences between the MC PDFs and their 
baseline 
counterparts shown in
Figs.~\ref{fig:xdep-q1geV-smallx}--\ref{fig:xdep-q100-impact-MCPDF-constraints}.

\begin{figure}[!tb]
  \includegraphics[width=\textwidth]{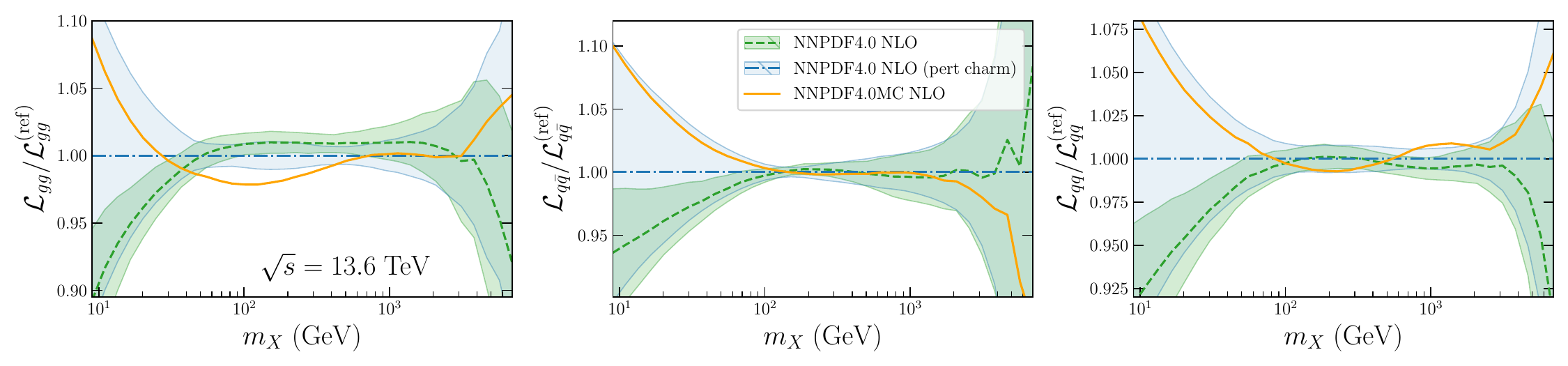}
  \caption{The gluon-gluon, quark-antiquark, and quark-quark parton
    luminosities  at the LHC with $\sqrt{s}=13.6$~TeV
  as a function of the invariant mass
  $m_X$ for the same PDFs as in
  Fig.~\ref{fig:xdep-q100-impact-MCPDF-constraints}, shown as a ratio
  to the NNPDF4.0 baseline.
  }
  \label{fig:lumi-mcpdfs.pdf}
\end{figure}

First, in
Fig.~\ref{fig:lumi-mcpdfs.pdf} we display the gluon-gluon,
quark-antiquark, and quark-quark parton luminosities at the LHC with $\sqrt{s}=13.6$~TeV
  as a function of the invariant mass of the final state $m_X$,
  computed from the same PDFs shown in
  Fig.~\ref{fig:xdep-q100-impact-MCPDF-constraints}, and shown as a
  ratio to the NNPDF4.0 baseline.
The luminosities are integrated over the full rapidity range  
and are thus dominated by the PDF behavior in the central rapidity
region, where $x_1\sim x_2\sim m_X/\sqrt{s}$. For   $50~{\rm GeV}\lsim
m_X \lsim 1~{\rm TeV}$  this is a
medium-small $x$ region, where
  differences between the MC PDFs and the baseline are generally
  moderate and only noticeable for the gluon. Indeed, in the case of
  the gluon-gluon luminosity MC PDFs lead to a   suppression of around
  2\%  in comparison to the baseline
for $100~{\rm GeV}\lsim m_X \lsim 3~{\rm TeV}$, while otherwise
differences between NNPDF4.0 NLO and its MC variant are at the 1\%
level, and only
become larger, though well within uncertainties,  for $m_X \lsim 100$~GeV  due to stronger small $x$ rise
of the MC PDFs.

\begin{figure}[!tb]
  \includegraphics[width=0.32\textwidth]{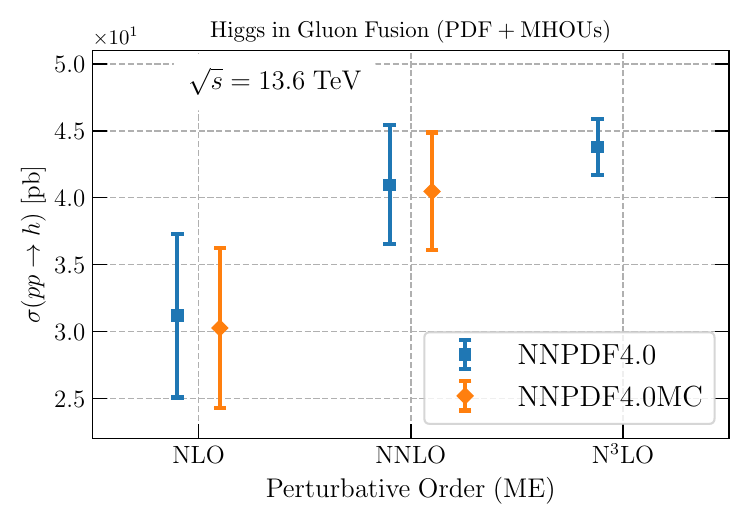}
  \includegraphics[width=0.32\textwidth]{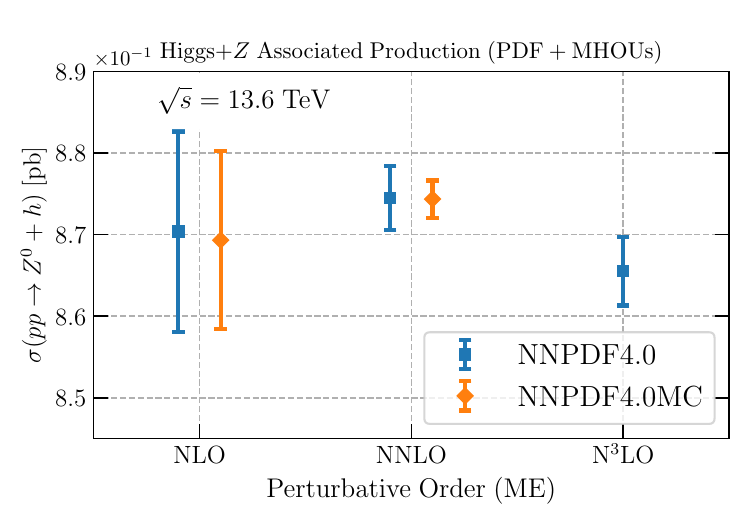}
  \includegraphics[width=0.32\textwidth]{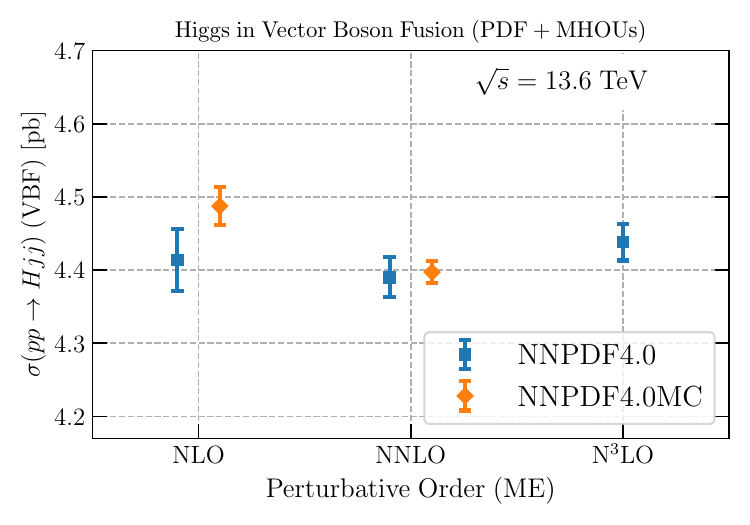}
  \includegraphics[width=0.32\textwidth]{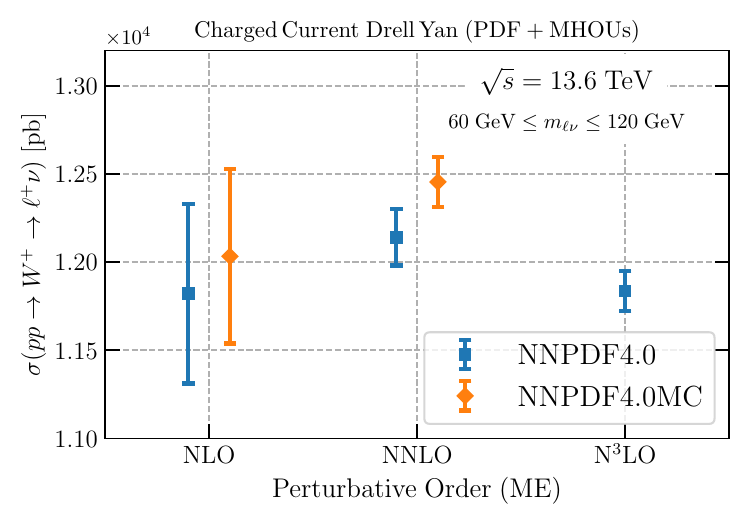}
  \includegraphics[width=0.32\textwidth]{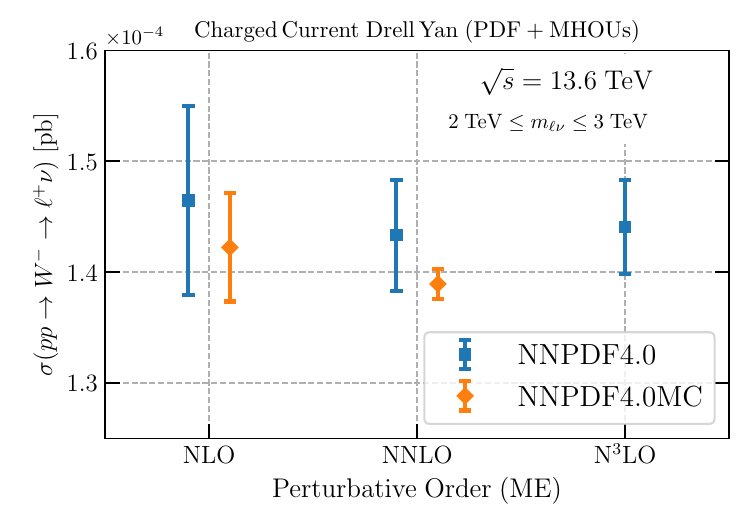}
  \includegraphics[width=0.32\textwidth]{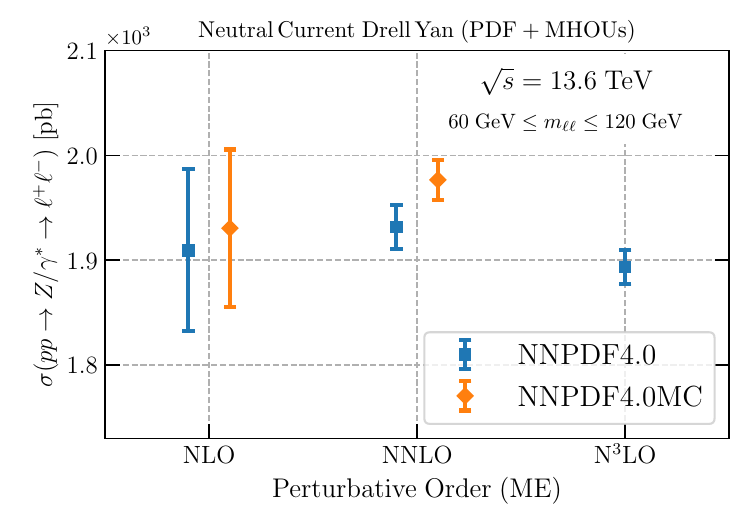}
  \caption{The inclusive NLO  and  NNLO cross-sections for Higgs production in gluon
    fusion, in association with a $Z$ boson, and in vector boson
    fusion (top), and on-shell and high-mass $W$ and on-shell $Z$
    production at the LHC $\sqrt{s}=13.6$~TeV (bottom), comparing
    NNPDF4.0MC PDFs and the baseline.  For the baseline NNPDF4.0, the
    aN$^3$LO result is also shown. The uncertainty shown is  scale
    variation with 7-point prescription only for the MC PDFs, combined
    in quadrature with the PDF uncertainty for the baseline sets.
  }
  \label{fig:lhc_inclusive_xsecs}
\end{figure}

We then consider  representative inclusive hard cross-sections:
Higgs and gauge boson production at the LHC with
$\sqrt{s}=13.6$~TeV, computed using  the {\sc\small ggHiggs}~\cite{Bonvini:2014jma}, {\sc\small n3loxs}~\cite{Baglio:2022wzu} and {\sc\small proVBFH}~\cite{Cacciari:2015jma,Dreyer:2016oyx} codes.
In Fig.~\ref{fig:lhc_inclusive_xsecs} we compare results obtained at
NLO and NNLO (both for PDF and the matrix element) with the MC sets
and their baseline counterparts, and for the latter also aN$^3$LO,
using  the settings
of Ref.~\cite{NNPDF:2024nan}. The uncertainty shown is for the MC sets
only that related to missing higher orders in the matrix element,
evaluated   from standard
7-point scale variation, while for the baseline sets it also includes 
the PDF uncertainty, combined in quadrature with it. The corresponding
uncertainty bands always overlap, reflecting the 
differences seen in parton luminosities. 

\begin{figure}[!tb]
\centering
  \includegraphics[width=0.44\textwidth]{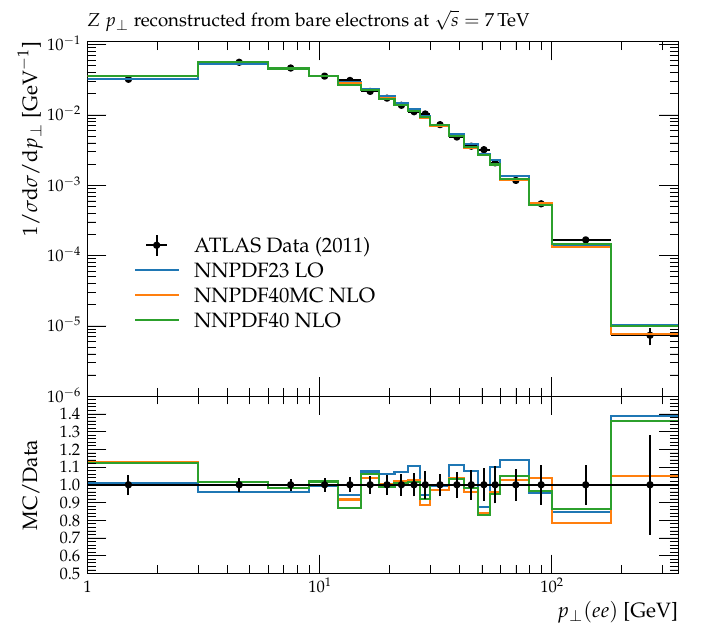}
  \includegraphics[width=0.44\textwidth]{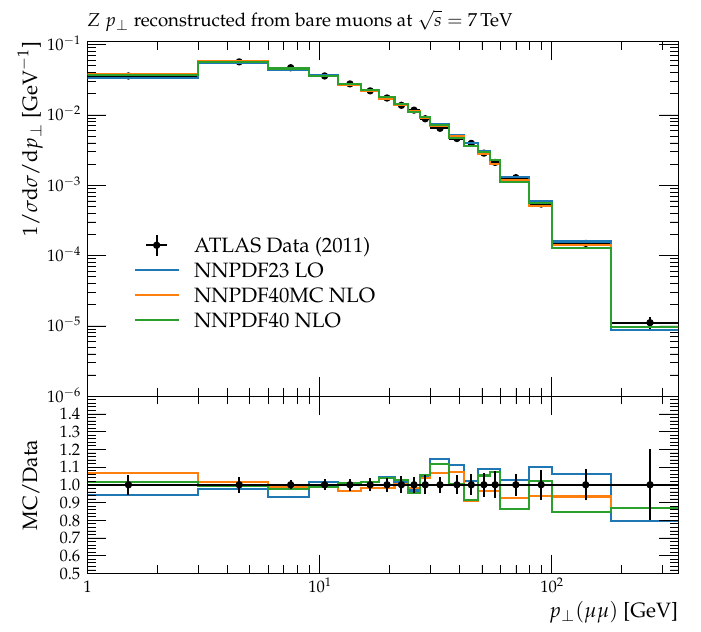}
  \caption{The normalized $Z$ boson $p_T$ distribution
    computed at LO using  {\sc\small Pythia8} and {\sc\small Rivet}
    using NNPDF2.3LO, NNPDF4.0 NLO,
and NNPDF4.0MC NLO PDFs. Predictions are compared to the  ATLAS~\cite{ATLAS:2011uxy} data
    at $\sqrt{s}=7~\mathrm{TeV}$ using bare electron (left) or muon
    (right) pairs; error bars on the data include statistical and systematic uncertainties.
    Both the absolute distribution (top) and the ratio of the theory
    prediction to the data (bottom) are shown.
  }
  \label{fig:zpt-distributions}
\end{figure}

We turn next to processes that are also sensitive to soft physics. We
show results for LHC differential distributions at
leading order  obtained from {\sc\small Pythia8} simulations
interfaced to the {\sc\small Rivet} analysis toolkit~\cite{Bierlich:2019rhm}.
We neglect PDF uncertainties and only display the central values found
using NNPDF2.3LO, NNPDF4.0 NLO,
and NNPDF4.0MC NLO PDFs.
We first consider the normalized $Z$ boson transverse momentum
distribution, reconstructed from bare dilepton events, either
electrons or muons, which is sensitive to both soft
and hard QCD. In Fig.~\ref{fig:zpt-distributions}
the {\sc\small Pythia8} LO predictions for  $1~{\rm GeV}\le
p_\perp(\ell\ell)\le 300~{\rm GeV}$ are compared to ATLAS data at
7~TeV from Ref.~\cite{ATLAS:2011uxy}. The low and high $p_\perp$
regions respectively probe  soft and hard QCD radiation.  
For the
normalized distributions shown, higher-order QCD corrections partially cancel
out. The difference between PDF sets is negligible, and  good
agreement with the data is found using all PDF sets except at very
small $p_\perp$ in the electron channel.

\begin{figure}[!tb]
\centering
    \includegraphics[width=0.44\textwidth]{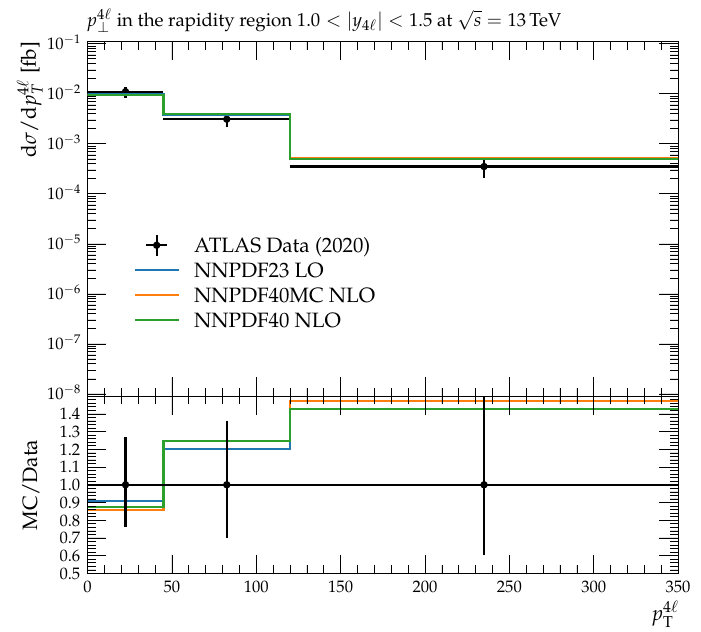}
    \includegraphics[width=0.44\textwidth]{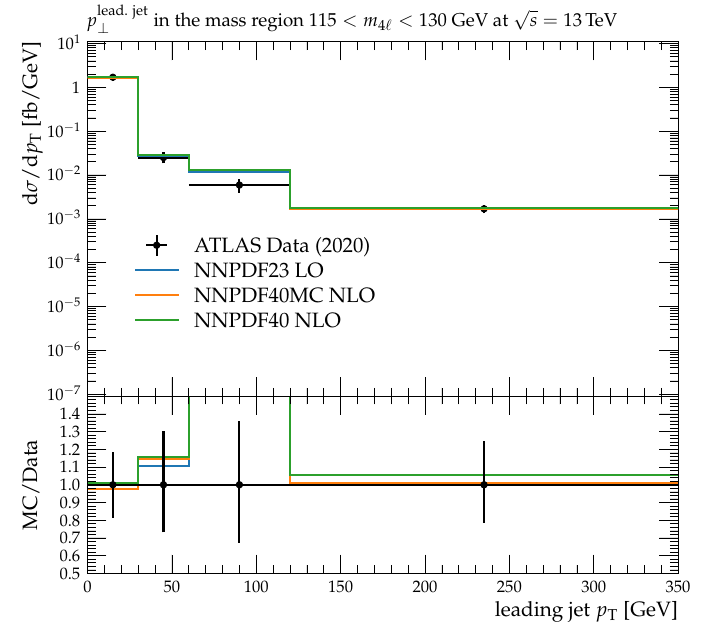}
    \caption{Same as Fig.~\ref{fig:zpt-distributions} for  the
      fiducial cross-section for Higgs production at $\sqrt{s} = 13~\mathrm{TeV}$ in the $H \to ZZ^\star \to 4 \ell ~(\ell = e, \mu)$
        decay channel. The four-lepton $p_T$ distribution
        for $1.0 < |y_{4\ell}| < 1.5$ (left) and the $p_T$ of the leading jet in events with $\ge 1$ jet  for
        $115 < m_{4\ell} < 130~\mathrm{GeV}$ (right) are shown,
        compared to ATLAS data~\cite{ATLAS:2020wny}.
    }
    \label{fig:pheno-higgs-ZZ-4leptons}
\end{figure}

We next consider  the 
fiducial cross-sections for Higgs production in the $H \to ZZ^\star \to 4 \ell ~(\ell = e, \mu)$
decay channel. In Fig.~\ref{fig:pheno-higgs-ZZ-4leptons} we compare
predictions to the ATLAS data collected at
$\sqrt{s} = 13~\mathrm{TeV}$ with an integrated luminosity of
$\mathcal{L}=139~\mathrm{fb}^{-1}$~\cite{ATLAS:2020wny}. Results are
shown for the transverse momentum distribution of the four hardest
leptons in the event, $p_T^{4\ell}$, in the rapidity range $1.0 < |y_{4\ell}| < 1.5$, and for the
transverse momentum of the leading jet in the invariant mass range 
$115 < m_{4\ell} < 130~\mathrm{GeV}$. Also in this case, differences
between different PDF sets are negligible, and  good agreement with
the data is found.

\begin{figure}[!tb]
\centering
    \includegraphics[width=0.44\textwidth]{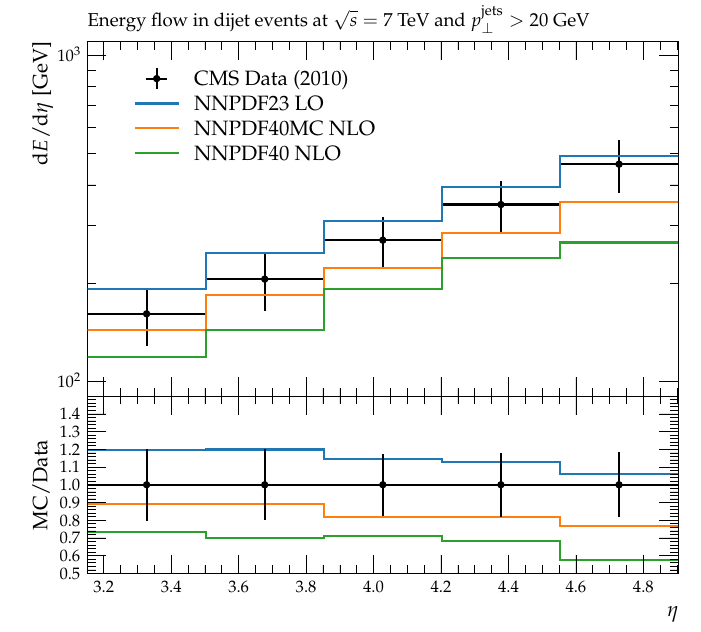}
    \includegraphics[width=0.44\textwidth]{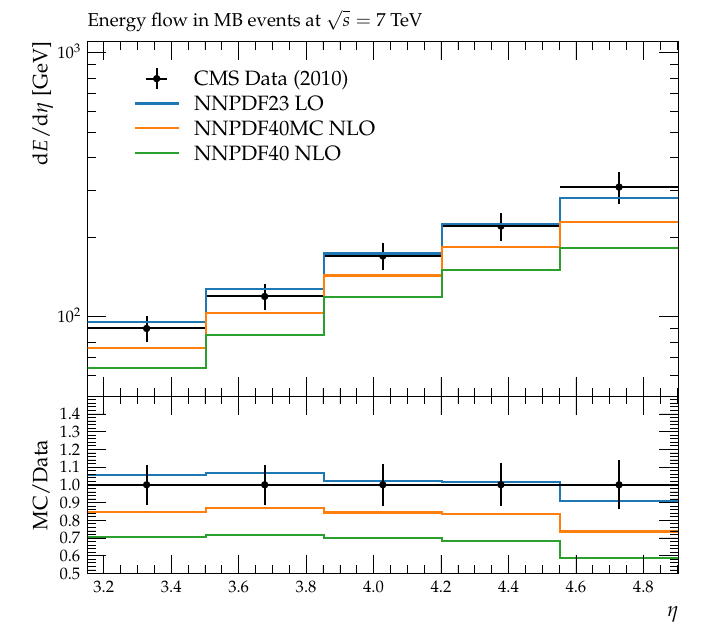}
    \caption{Same as Fig.~\ref{fig:zpt-distributions}, now for the
      energy flow in dijet (left) and minimum-bias  events with
      $\sqrt{s}=7~\mathrm{TeV}$ and  $3.2 \le \eta \le 4.9$ compared to CMS data~\cite{CMS:2011xjg}.
    }
    \label{fig:pheno-energy-flow}
\end{figure}

We then turn to the energy flow, defined as
\be
\frac{dE}{d\eta} = \frac{1}{|\eta_{\rm max}-\eta_{\rm min}|} \lp \frac{1}{N_{\rm inel}} \sum_{i=1}^{n_{\rm part}} E_i \theta(\eta_i > \eta_{\rm min})
\theta(\eta_i < \eta_{\rm max}) \rp \, ,
\ee
where $\eta$ is the midpoint of the rapidity interval, $\lc \eta_{\rm
  min},\eta_{\rm max}\rc$, $N_{\rm inel}$ is the number of inelastic pp
collisions, and $n_{\rm part}$ is the number of stable particles in the
event whose energy is equal to $E_i$.
The energy flow in dijet events and in minimum-bias  events  at
$\sqrt{s}=7~\mathrm{TeV}$ in the forward  $3.2 \le \eta \le 4.9$
ranges is shown
in Fig.~\ref{fig:pheno-energy-flow}, compared to the CMS data of~\cite{CMS:2011xjg}.
For the dijet sample, a $p_\perp^{\rm jet}>20$ GeV cut is imposed.
For both dijet and minimum-bias events, the simulations based on NNPDF2.3LO 
display good agreement with the data, while those obtained using
NNPDF4.0 NLO sets (both MC and baseline)
tend to undershoot the experimental measurements, which suggests the need for a dedicated tune of soft QCD physics.

\begin{figure}[!tb]
\centering
  \includegraphics[width=0.44\textwidth]{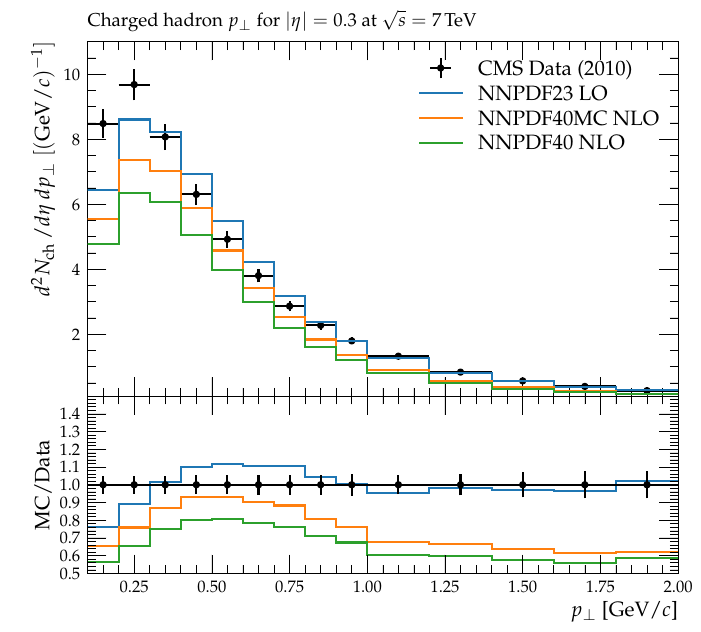}
  \includegraphics[width=0.44\textwidth]{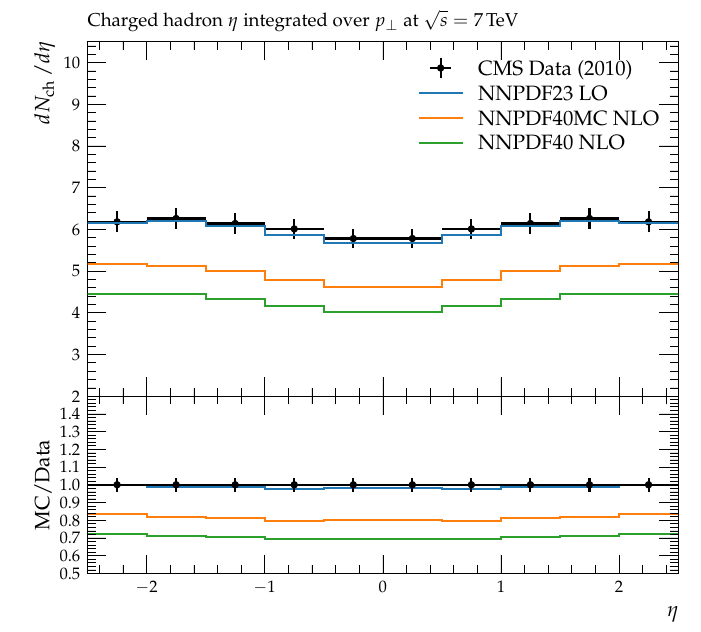}
  \caption{Same as Fig.~\ref{fig:pheno-energy-flow}, now for the charged-hadron transverse momentum (left) and pseudorapidity (right) distributions in proton-proton collisions at 
    $\sqrt{s}=7~\mathrm{TeV}$, comparing to the CMS measurements of~\cite{CMS:2010tjh}.
  }
  \label{fig:pheno-charged-distributions}
\end{figure}

Finally, in Fig.~\ref{fig:pheno-charged-distributions} we show the
charged-hadron multiplicity distribution, differential in pseudorapidity
and in transverse momentum, $d^2 N_{\rm ch} / d\eta dp_\perp$, as a function of $p_\perp$ at fixed rapidity $\lvert \eta \lvert = 0.3$
and as a function of $\eta$  integrated over the full $p_\perp$
range. Predictions are 
 compared to the CMS measurements of~\cite{CMS:2010tjh}, for  events
that satisfy both $p_\perp \leq 2~\mathrm{GeV}$ and $\lvert \eta
\lvert < 2.5$ in order to highlight the sensitivity to the modeling
of nonperturbative QCD dynamics.
As in the case of the energy flow, the NNPDF2.3LO set provides the best
description of the experimental data, while the NNPDF4.0 sets
undershoot the CMS measurements.
Indeed, both the energy flow of Fig.~\ref{fig:pheno-energy-flow} and
the charged-hadron differential distributions of
Fig.~\ref{fig:pheno-charged-distributions} are sensitive to
non-perturbative QCD processes. It follows that achieving a good description
requires a dedicated tune of soft QCD, and differences seen in
Figs.~\ref{fig:pheno-energy-flow}-\ref{fig:pheno-charged-distributions}
do not have a simple physical interpretation, and are simply a
manifestation of the fact that the NNPDF4.0 sets have not been used in
the Monte Carlo tune.
The  Monash 2013 tune of
{\sc\small Pythia8} used here is based on NNPDF2.3LO, explaining
the good agreement found for this set. 

\section{Summary and outlook}
\label{sec:summary}

The NNPDF4.0MC PDFs presented in this work satisfy the requirements of
event generators not only at LO but also at NLO and
NNLO accuracy, while the NLO and NNLO sets  provide a
satisfactory description of the global dataset and minimize
differences in comparison to the baseline sets, ensuring their
reliability to evaluate hard cross-sections at the LHC and
elsewhere.
It thus  becomes possible to combine the precision and accuracy
enjoyed by global PDF sets at NLO and NNLO without compromising the
usability of these PDFs in  generators for initial-state radiation
and the modeling of soft QCD processes.

In order to also achieve
agreement with the data  for non-perturbative
processes such as the underlying event, pileup, and low-$p_T$ radiation, the
soft QCD models specific to each event generator will need to be tuned to
the data using as input these new NNPDF4.0MC PDFs, since their behavior,
especially for low-$x$ physics, becomes a component of the
tuning model. 
Such dedicated tunes will be needed in order for the NNPDF4.0MC PDF sets
to become instrumental in the development of a next generation of
Monte Carlo codes that reaches higher perturbative accuracy. To this
purpose, we aim  to collaborate with event generator developers in
order to
integrate NNPDF4.0MC in their frameworks and produce dedicated tunes
of soft QCD physics such that the whole palette of LHC processes, from
the soft to the perturbative region, can be satisfactory described
within a single 
physics simulation.

The NNPDF4.0 MC sets are made available through the {\sc\small LHAPDF} interface~\cite{Buckley:2014ana} and  the NNPDF Collaboration website.\footnote{\url{https://nnpdf.mi.infn.it/nnpdf4-0-mc/}}

\subsection*{Acknowledgments}

We are very grateful to Gavin Salam and Peter Skands for many useful discussions,
productive feedback, and extensive benchmarking concerning the MC PDF sets presented in this work.
We also thank Melissa van Beekveld, Silvia Ferrario Ravasio,
Christian Gutschow, Max Knobbe, Frank Krauss, and Oliver Mattelaer for assistance with
the validation of the NNPDF4.0 MC PDFs.
J.~R. and T.~R. are partially supported
by NWO, the Dutch Research Council, and by the Netherlands eScience Center (NLeSC). 

\bibliography{nnpdf40_tuning}

\end{document}